\documentclass[times, 10pt,twocolumn]{IEEEconf} 

\newfont{\mycrnotice}{ptmr8t at 7pt}
\newfont{\myconfname}{ptmri8t at 7pt}

\usepackage[hyphens]{url}
\usepackage{hyperref}
\usepackage{listings}
\usepackage{graphicx}
\usepackage{amstext}
\usepackage{bbm}
\usepackage{amsmath}
\usepackage{amsfonts}
\usepackage{amssymb}
\usepackage{amsthm}
\usepackage{xspace}
\usepackage{multirow}

\usepackage{color}

\usepackage{epstopdf}

\renewcommand{\url}{\path}

\newcommand{\eg}{e.g.\xspace}

\newcommand{\apriori}{\emph{a priori}\xspace}

\newcommand{\vect}[1]{\ensuremath{\mathbf{#1}}}

\DeclareMathOperator{\pdf}{p}
\DeclareMathOperator{\Pb}{P}

\begin{document}

\title{Deep Neural Network Based Malware Detection Using Two Dimensional Binary Program Features}

\author{Joshua Saxe\thanks{Authors contributed equally to the work.}\\
	\begin{affiliation}
		Invincea Labs, LLC 
	\end{affiliation}\\
	\email{josh.saxe@invincea.com}
\and
Konstantin Berlin\footnotemark[1]\\
	\begin{affiliation}
		Invincea Labs, LLC 
	\end{affiliation}\\
	\email{kberlin@invincea.com}
}

\maketitle

\begin{abstract}
Malware remains a serious problem for corporations, government agencies, and individuals, as attackers continue to use it as a tool to effect frequent and costly network intrusions.  Today malware detection is still done mainly with heuristic and signature-based methods that struggle to keep up with malware evolution.  Machine learning holds the promise of automating the work required to detect newly discovered malware families, and could potentially learn generalizations about malware and benign software (benignware) that support the detection of entirely new, unknown malware families.  Unfortunately, few proposed machine learning based malware detection methods have achieved the low false positive rates and high scalability required to deliver deployable detectors.

In this paper we introduce an approach that addresses these issues, describing in reproducible detail the deep neural network based malware detection system that Invincea has developed.  Our system achieves a usable detection rate at an extremely low false positive rate and scales to real world training example volumes on commodity hardware.  Specifically, we show that our system achieves a  95\% detection rate at 0.1\% false positive rate (FPR), based on more than 400,000 software binaries sourced directly from our customers and internal malware databases. We achieve these results by directly learning on all binaries, without any filtering, unpacking, or manually separating binary files into categories. Further, we confirm our false positive rates directly on a live stream of files coming in from Invincea's deployed endpoint solution, provide an estimate of how many new binary files we expected to see a day on an enterprise network, and describe how that relates to the false positive rate and translates into an intuitive threat score.

Our results demonstrate that it is now feasible to quickly train and deploy a low resource, highly accurate machine learning classification model, with false positive rates that approach traditional labor intensive signature based methods, while also detecting previously unseen malware.  Since machine learning models tend to improve with larger data-sizes, we foresee deep neural network classification models gaining in importance as part of a layered network defense strategy in coming years.

\end{abstract}

\section{Introduction}

Malware continues to facilitate crime, espionage, and other unwanted activities on our computer networks, as attackers use malware as a key tool their campaigns .  One problem in computer security is therefore to detect malware, so that it can be stopped before it can achieve its objectives, or at least so that it can be expunged once it has been discovered.

In this vein, various categories of detection approaches have been proposed, including, on the one hand, rule or signature based approaches, which require analysts to hand craft rules that reason over relevant data to make detections, and, on the other hand, machine learning approaches, which automatically reason about malicious and benign data to fit detection model parameters.  A middle path between these two methods is the automatic generation of signatures.  To date, the computer security industry, to our knowledge, has favored manual and automatically created rules and signatures over machine learning and statistical methods, because of the low false positive rates achievable by rule and signature-based methods.   

In recent years, however, a confluence of three developments have increased the possibility for success in machine-learning based approaches, holding the promise that these methods might achieve high detection rates at low false positive rates without the burden of human signature generation required by manual methods.

The first of these trends is the rise of commercial threat intelligence feeds that provide large volumes of new malware, meaning that for the first time, timely, labeled malware data are available to the security community.  The second trend is that computing power has become cheaper, meaning that researchers can more rapidly iterate on malware detection machine learning models and fit larger and more complex models to data.  Third, machine learning as a discipline has evolved, meaning that researchers have more tools at their disposal to craft detection models that achieve breakthrough performance in terms of both accuracy and scalability.

In this paper we introduce an approach that takes advantage of all three of these trends: a deployable deep neural network based malware detector using static features that gives what we believe to be the best reported accuracy results of any previously published detection engine.

The structure of the rest of this paper is as follows.  In Section \ref{sec:methods} we describe our approach, giving a description of our feature extraction methods, our deep neural network, and our Bayesian calibration model.  In Section \ref{sec:evaluation} we provide multiple validations of our approach.  Section \ref{sec:background} treats related work, surveying relevant malware detection research and comparing our results to other proposed methods.  Finally, Section \ref{sec:conclusion} concludes the paper, reiterating our findings and discussing plans for future work.

\section{Method}
\label{sec:methods}

\begin{figure}[ht]
	\center
	\includegraphics[width=0.45\textwidth]{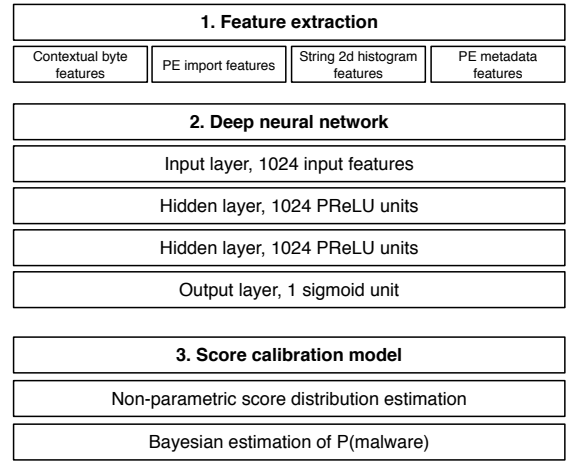}
   \caption{Outline of our malware classification framework.}
   	\label{fig:overview}
\end{figure}

Our full classification framework, shown in Fig. \ref{fig:overview}, consists of three main components.  The first component extracts four different types of complementary features from the static benign and malicious binaries.  The second component is our deep neural network classifier which consists of an input layer, two hidden layers and an output layer.  The final component is our score calibrator, which translates the outputs of the neural network to a score that can be realistically interpreted as approximating the probability that the file is actually malware.  In the remainder of this section we describe each of these model components in detail.

\subsection{Feature Engineering}
\label{sec:features}

\subsubsection{Byte/Entropy Histogram Features}

The first set of features that we compute for input binaries are the bin values of a two-dimensional byte entropy histogram that models the file's distribution of bytes. 

To extract the byte entropy histogram, we slide a 1024 byte window over an input binary, with a step size of 256 bytes. For each window, we compute the base-2 entropy of the window, and each individual byte occurrence in the window  (1024 non-unique values) with this computed entropy value, storing the 1024 pairs in a list.

Finally, we compute a two-dimensional histogram over the pair list, where the histogram entropy axis has sixteen evenly sized bins over the range $[0,8]$, and the byte axis has sixteen evenly sized bins over the range $[0,255]$. To obtain a feature vector (rather than the matrix), we concatenate each row vector in this histogram into a single, 256-value vector.

Our intuition in using these features is to model the contents of input files in a file-format agnostic way.  We have found that in practice, the effect of representing byte values in the entropy ``context'' in which they occur separates byte values that occur in the context of, for example, x86 instruction data from, for example, byte values occurring in compressed data.

\subsubsection{PE Import Features}

The second set of features that we compute for input binaries are derived from the input binary's import address table.  We initialize an array of 256 integers to zero; extract the import address table from the binary program; hash each tuple of DLL name and import function into the range $[0,255]$; and increment the associated counter in our feature array.  

Our intuition is that import table DLLs may help our model to capture the semantics of the external function calls that a given input binary relies upon, thereby detecting either heuristically suspicious files or files with a combination of imports that match a known malware family.  By hashing the potentially large number of imported functions into a small array, we ensure that our feature space remains fixed-size, which is important for scalability.  In practice we find that even with a 256 valued hash function  our neural network learns a meaningful separation between malware and benignware, as shown later in our evaluation.

\subsubsection{PE Metadata Features}

The final set of features are derived from the numerical fields extracted from target binary's portable executable (PE) packaging.  The portable executable format is the standard format for executables on Windows-family operating systems.  To extract these features we extract numerical portable executable fields from the binary using the using ``pefile'' Python parsing library.  Each of these fields has a textual name (\eg, ``compile\_timestamp"), which, similar to the import table, we aggregate into 256-length array. 

Our motivation for extracting these features is to give our model the opportunity to identify both heuristically suspicious aspects of a given binary program's packaging, and allow it to learn signatures that capture individual malware families.

\subsubsection{Aggregation of Feature Types}

To construct our final feature vector, we take the four 256-dimensional feature vectors described above and concatenate them into a single, 1024-dimensional feature vector. We found, throughout the course of our research, that this reduction of data into \apriori fixed sized small vector resulted in only a minor degradation in the accuracy of our model, and allowed us to dramatically reduce the memory and CPU time necessary to load and train our model, as compared to the more common approach of assigning each categorical value to its own column in the feature vector.

\subsubsection{Labeling}

To train and evaluate our model at low false positive rates, we require accurate labels for our malware and benignware binaries.  We accomplish this by running all of our data through VirusTotal, which runs the binaries through approximately 55 malware engines.We then use a voting strategy to decide if each file is either malware or benignware. 

Similar to \cite{berlin2015detecting}, we label any file against which 30\% or more of the anti-virus engines alarm as malware, and any file that no anti-virus engine alarms on as benignware.  For the purposes of both training and accuracy evaluation we discard any files that more than 0\% and less than 30\% of VirusTotal's anti-virus engines declare it malware, given the uncertainty surrounding the nature of these files. Note that we do not filter our binary files based on any actual content, as this could bias our results.

\subsection{Neural Network}

For classification, we use a deep feedforward neural network consisting of four layers, where the first three 1024 node layers consist of a dropout \cite{srivastava2014dropout}, followed by a dense layer with either, a parametric rectified linear unit (PReLU) activation function \cite{he2015delving} in the first two layers, or the sigmoid function, in the last hidden layer (the fourth layer being the prediction). We elaborate on the reasoning behind these choices below.

\subsubsection{Design}

First, our choice of using deep neural network, rather than a shallow but wide neural network, is based on the developed understanding that deep architectures can be more efficient (in terms of number of fitting parameters) than shallow network \cite{bengio2007scaling}. This is important in our case, since the number of binary samples in our dataset is still relatively small, as compared to all the possible binaries that can observed on a large enterprise network, and so our sampling of the feature space is limited. Our goal was to increase expressiveness of the network, while maintaining a tractable size network that can be quickly trained on a single Amazon EC2 node. Given our four layer neural network design, the remaining design choices are meant to address overfitting and improve efficacy of the backpropagation algorithm.

\subsubsection{Preventing Overfitting}

Dropout has been demonstrated to be a very simple and efficient approach for preventing overfitting in deep neural network. Unlike standard weight regularizers, such as based on the $\ell_1$ or $\ell_2$ norms, that push the weights toward some expected prior distribution \cite{friedman2001elements}, dropout seeks weights at each node that are complementary to weights in other nodes. The dropout solution is potentially more resilient to imperfect or dirty data (which is common when extracting features from similar malware that was compiled or packed using different software), since it discourages co-adaptation by creating multiple paths to correct classification throughout the network. This can be viewed as implicit bagging of several neural network models \cite{srivastava2014dropout}.

\subsubsection{Speeding Up Learning}

Rectified linear units (ReLU) have been shown to significantly speedup network training over traditional sigmoidal activation functions, such as $\tanh$ \cite{krizhevsky2012imagenet}, by avoiding significant decay in gradient descent convergence rate after an initial set of iterations. This slowdown is due to saturating non-linearities in sigmoidal functions at their edges \cite{krizhevsky2012imagenet, maas2013rectifier,he2015delving}. Using ReLU activation functions can also lead to bad performance when the input values are below 0, and PReLU activator functions are made to dynamically adjust in order to avoid this issue, thus yielding significantly improved convergence rate \cite{he2015delving}.

Initialization of weights, before training, can significantly impact the convergence of the backpropagation algorithm \cite{glorot2010understanding,he2015delving}. The goal of a good initialization is to avoid multiplicative impact of weight aggregation from multiple layers during backpropagation. In our approach we use the Gaussian distribution that is normalized based on the size of the input and output of the layers, as suggested in \cite{glorot2010understanding}.  Before doing this initialization we transform our feature values by applying the base-10 logarithm to each feature value, which we found in practice improved training performance substantially.

\subsubsection{Formal Description}

Let $l=\{0,1,2,3\}$ be a layer in the network, $\vect{y}^{(l-1)}$ the incoming values into the layer (for $l=1$ those are the feature values), $\vect{y}^{(l)}$ the output values of the layer, $\vect{W}^{(l)}$ the weights of the layer that linearly transforms $n$ input values into $m$ output values, $b^{l}$ the bias, and $F^{(l)}$ the associated activation vector function. The equation for $l=\{1,2,3\}$ of the network is
\begin{equation}
\begin{aligned}
	\vect{d}^{(l)} &= \vect{y}^{(l-1)} \cdot \vect{r}^{(l)},\\
	\vect{z}^{(l)} &= \vect{W}^{(l)} \vect{d}^{(l)} +b^{(l)},\\
	\vect{y}^{(l)} & = F(\vect{z}^{(l)}),
\end{aligned}
\end{equation}
where $\cdot$ is a pointwise (elementwise) vector product, and $r_i$ are independent samples from a Bernoulli distribution with parameter $h$. The $\vect{r}$ values are resampled for each batch update during training, and $h$ corresponds to the fraction of nodes that are kept during each batch update \cite{srivastava2014dropout}. Layer $l=0$ is the input layer, and $l=4$ is the output layer.

For layers $l=\{1,2\}$, the activation function is the PReLU function, 
\begin{equation}
	F(\vect{z}_i^{(l)}) = (y_1^{(l)},\ldots, y_i^{(l)}, \ldots, y_m^{(l)})
\end{equation}
where for some additional parameter $a_i^{(l)}$ that is also fit during training,
\begin{equation}
\begin{aligned}
     y_i^{(l)}=
	  \begin{cases}
				a_i^{(l)} z_i^{(l)} & \text{if $z_i^{(l)}<0$},\\
				z_i^{(l)} & \text{else.}	
			\end{cases}
\end{aligned}
\end{equation}
For the final layer $l=3$, the activation function is the sigmoid function,
\begin{equation}
	y^{*}=\frac{1}{1+e^{-z^{(3)}}},
	\label{eq:sig}
\end{equation}
which produces the output of our model.

The loss for each $n$ sized batch sample is evaluated as the sum of the cross-entropy between the neural net's prediction and the true label,
\begin{equation}
	L(\vect{y}^{*}, \hat{\vect{y}}) = -\sum_{j=1}^{n}   \left[ \hat{y}_j \log y_j^{*} + (1-\hat{y}_j) \log (1- y_j^{*}) \right]
\end{equation}
where $\vect{y}^{*}$ is the output of our model for all $n$ batch samples, $y_j^{*}$ is the output for sample $j$, and $\hat{y}_j \in \{0,1\}$ is the true label of the sample $j$, with $0$ representing benignware and $1$ malware.

The neural network is training using backpropagation and the Adam gradient-based optimizer \cite{kingma2014adam}, which we observed to converge significantly faster than the standard stochastic gradient descent.

\subsection{Bayesian Calibration}

Beyond simply detecting malware in a binary sense our system also seeks to provide users with accurate probabilities that a given file is malware.  We do this through a Bayesian model calibration approach which combines our empirical belief about the "riskiness" of a given customer network (represented as our belief about the ratio of malware to benignware on the customer's network) and empirical information about our neural network's error profile against test data.  Here describe our specific approach for adjusting the classifier's ``probability'' score to reflect the true ``threat'' score, given this qualitatively assumed ratio of malware to benignware. 

Let $0\leq s \leq1$ be some score given by the classifier, reflecting the degree to which a classifier believes an observed binary is malware, with $0$ being completely benign, and $1$ being certainly malware. Our goal is to translate this number into a ``threat'' score, which will give the user a measure of how likely that the observed binary is actually malware. In line with this intuition, we define the threat score as the probability that the file will actually be malware, $\Pb(C=m | S=s)$, given the score $s$, and category $C=\{m,b\}$. We will use capital $\Pb$ for probability, and the little $\pdf$ to represent probability density function (pdf), and for brevity drop the equality sign.

Lets assume we have a pdfs for the benign and malware scores for a given classifier, $\pdf(S=s | C=m)$ and  $\pdf(S=s | C=b)$. We will describe in the next section how we derive these pdfs from observed test data. Given a base rate $r$, the ratio of malware to benignware, we will not derive how to compute the threat score. This will be done in two steps: i) we express our problem in terms of our classifier's expected pdf for benign and malware predictions, and ii) we demonstrate how to practically compute these pdfs.

\subsubsection{Threat Score}

Using Bayes' rule we have
\begin{equation}
	\Pb(m | s) = \frac{p(s | m)\Pb(m)}{p(s)}.
\end{equation}
Rewriting $\pdf(s)$ as the sum of probabilities over the two possible labels, we get
\begin{equation}
	\Pb(m | s) = \frac{\pdf(s | m)\Pb(m)}{\pdf(s | m)\Pb(m)+\pdf(s | b)\Pb(b)}.
\end{equation}

Finally, using the constraint that probabilities add up to $1$, gives us the final value of the threat score in terms of pdfs and probability of observing malware (malware base rate) ,
\begin{equation}
		\Pb(m | s)=  \frac{p(s | m)\Pb(m)}{\pdf(s | m)\Pb(m)+\pdf(s | b)(1-\Pb(m))}.
		\label{eq:threat}
\end{equation}

\subsubsection{Probability Density Function Estimation}

Given the above definition of the threat score, we need to derive the pdfs $\pdf(s | m)$ and $\pdf(s | b)$. There are two approaches that are commonly used: i) the parametric approach, where we assume some distribution for the pdfs, and fit the parameters of that distribution based on the observed samples, and ii) the non-parametric approach, like kernel density estimator (KDE), where we approximation a value of pdf given $C$ by taking a weighted average of the neighborhood. 

Since it is not reasonable to expect the output of our ML classifier to follow some standard distribution, we used KDE with the Epanechnikov kernel \cite{epanechnikov1969non}. In our testing it had better validation score than the standard Gaussian kernel. Since the pdfs can only take values in $[0,1]$, we mirrored our samples to the left of 0 and the right of 1, before computing the estimated pdf value at a specific point. The window size was set empirically to $0.01$ to better approximate the tail end of distributions, were samples are less dense.

\section{Evaluation}
\label{sec:evaluation}

We evaluated our system in two ways.  First, we used our in-house database of malicious and benign binaries to conduct a set of cross-validation experiments testing how well our system performs using the individual feature sets described above and the agglomeration of the feature sets described above.  Second, we used a live feed of binaries from Invincea customer networks, in conjunction with a live feed of malicious binaries from the Jotti subscription threat intelligence feed, to measure the accuracy of our system in deployment contexts using all feature sets.  

All our experiments were ran on Amazon EC2 g2.8xlarge instance, which has 60GB of RAM, and four 1,536 CUDA core graphical processing units, of which we only used one. The software uses the Keras v0.1.1 deep learning library to implement the neural network model described above. The feature extraction is mostly written in Cython and Python, heavily relying on SciPy and NumPy libraries, and each sample's features are extracted by a single thread process. Below we describe each of these evaluations in detail, starting with a description of our evaluation datasets and then moving on to descriptions of our methodology and results.

\subsection{Dataset}

\begin{figure}[h]
	\center
	\includegraphics[width=0.45\textwidth]{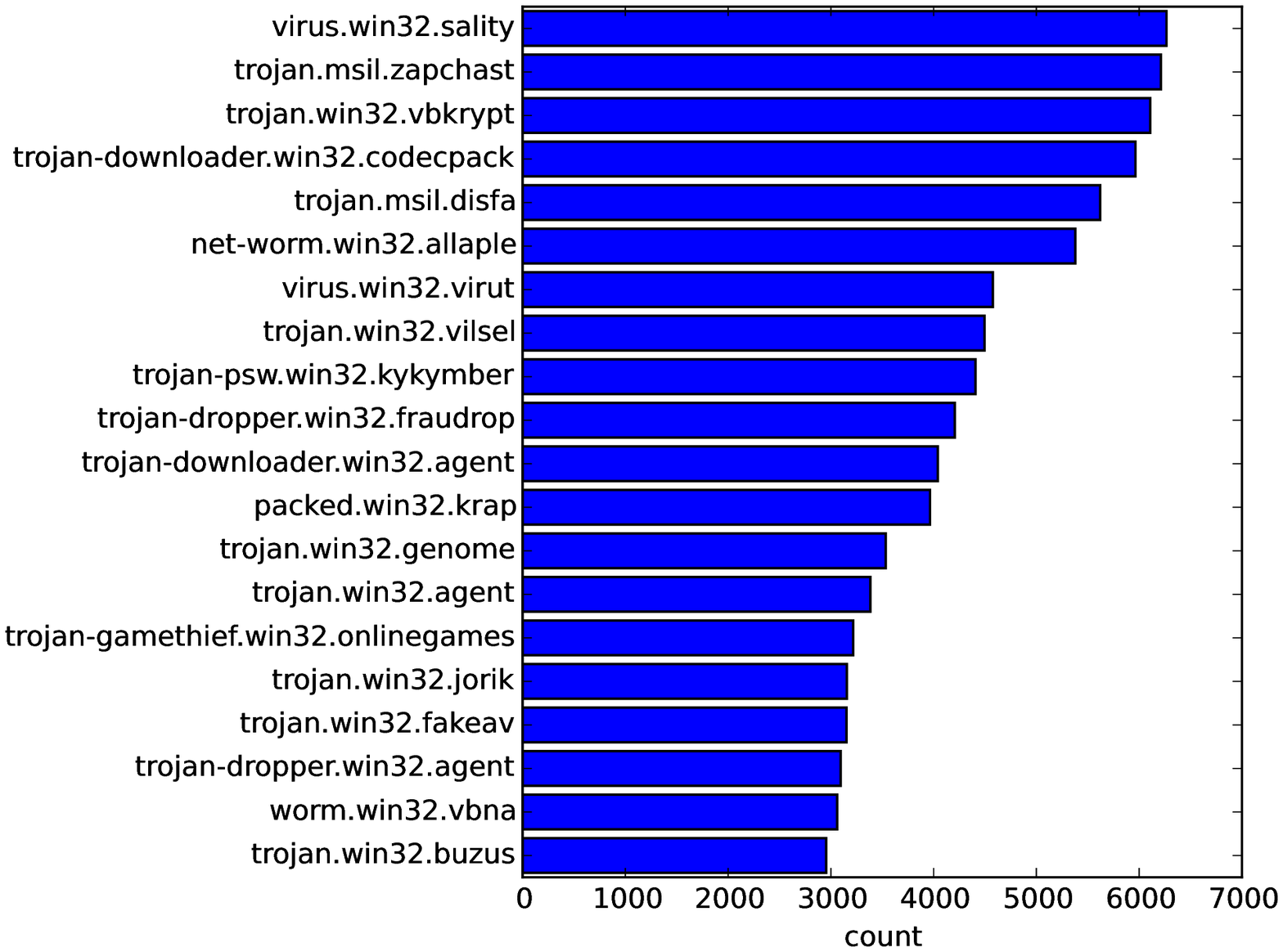}
   \caption{Counts of the top 20 malware families in our experimental dataset.}
   	\label{fig:families}
\end{figure}

\begin{figure}[h]
	\center
	\includegraphics[width=0.45\textwidth]{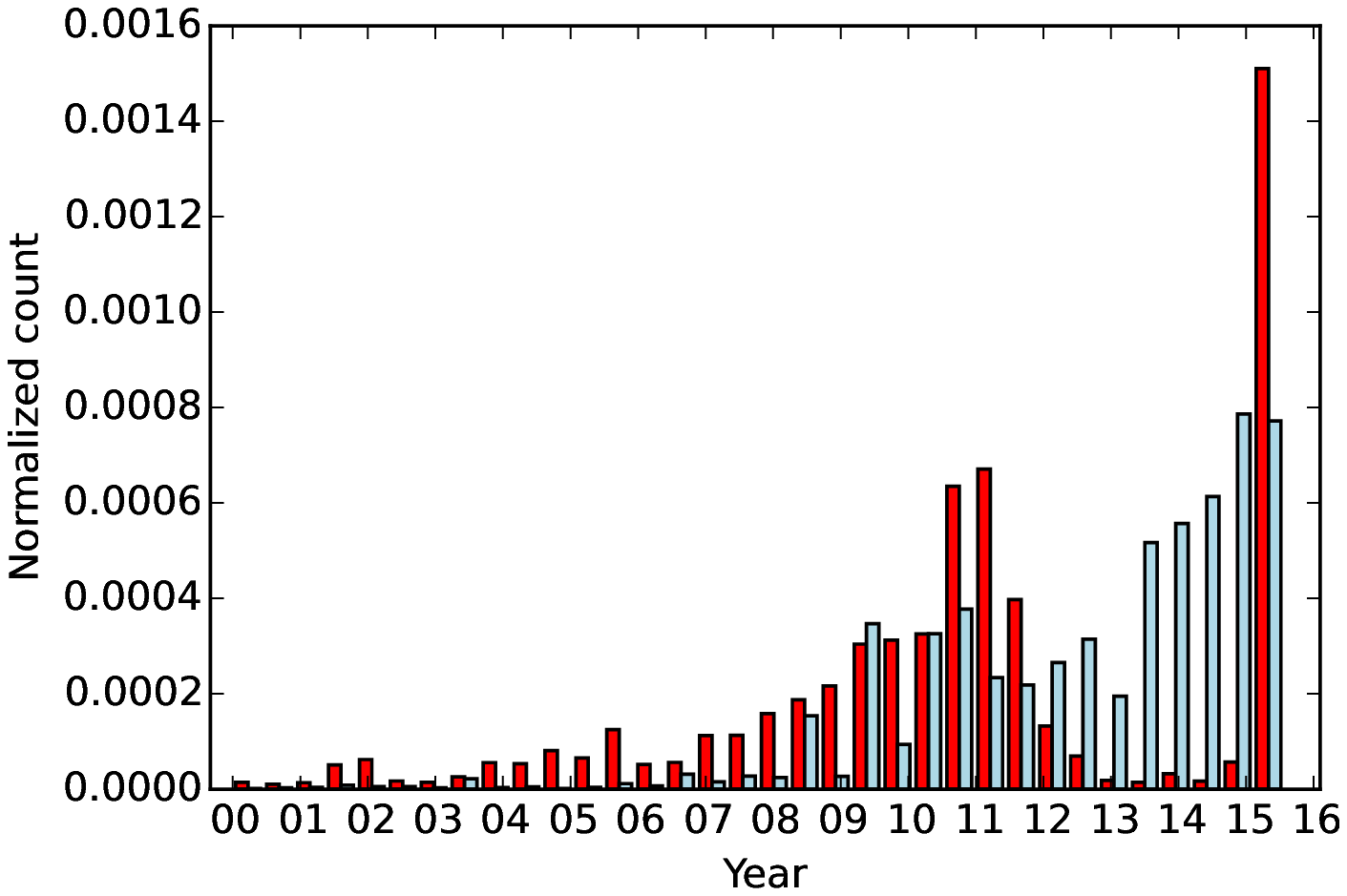}
   \caption{Normalized histogram of compile timestamps for our malware (left, red) and benignware (right, teal) datasets based on the Portable Executable compile timestamp field.}
   	\label{fig:bintimehist}
\end{figure}

Our benign and malware binaries were drawn from Invincea's own computer systems and Invincea's customers networks.  We used malicious binaries obtained from both the Jotti commercial malware feed and from Invincea's private malware database. Our final dataset, after VirusTotal filtering, contains 431,926 binaries, with 81,910 labeled as benignware and 350,016 as malware.  Fig. \ref{fig:families} shows counts for the top malware families, as identified by the Kaspersky anti-virus engine, in our malware dataset.  Fig. \ref{fig:bintimehist} gives a histogram over the compile timestamps of both the malicious and benign binaries in our dataset.  

Not all malware binaries discovered by the security community are shared publicly, or as part of threat intelligence feeds, and the distribution of malware binaries that a specific enterprise network might experience could differ somewhat from our dataset. However, it significantly more critical is to have an accurate distribution of the benign files that reflects real usage, since that is what will drive the critical FPR estimation. Since the primary source of benign data comes directly from Invincea's clients, we believe it is one the most accurate representations of the true distribution that has been evaluation. In our validation we confirm that our ROC curve FPR estimates match closely the live stream FPR (under the assumption that Invincea's endpoint stream contains little to no malware). 

\subsection{Cross-Validation Experiment}

We conducted five separate 4-fold cross-validation experiments, where for each experiment we randomly split our data into four equally sized partitions. For each of the four partitions, we trained against three partitions and tested against the fourth.

\begin{figure*}[ht]
	\center
	\includegraphics[width=1.0\textwidth]{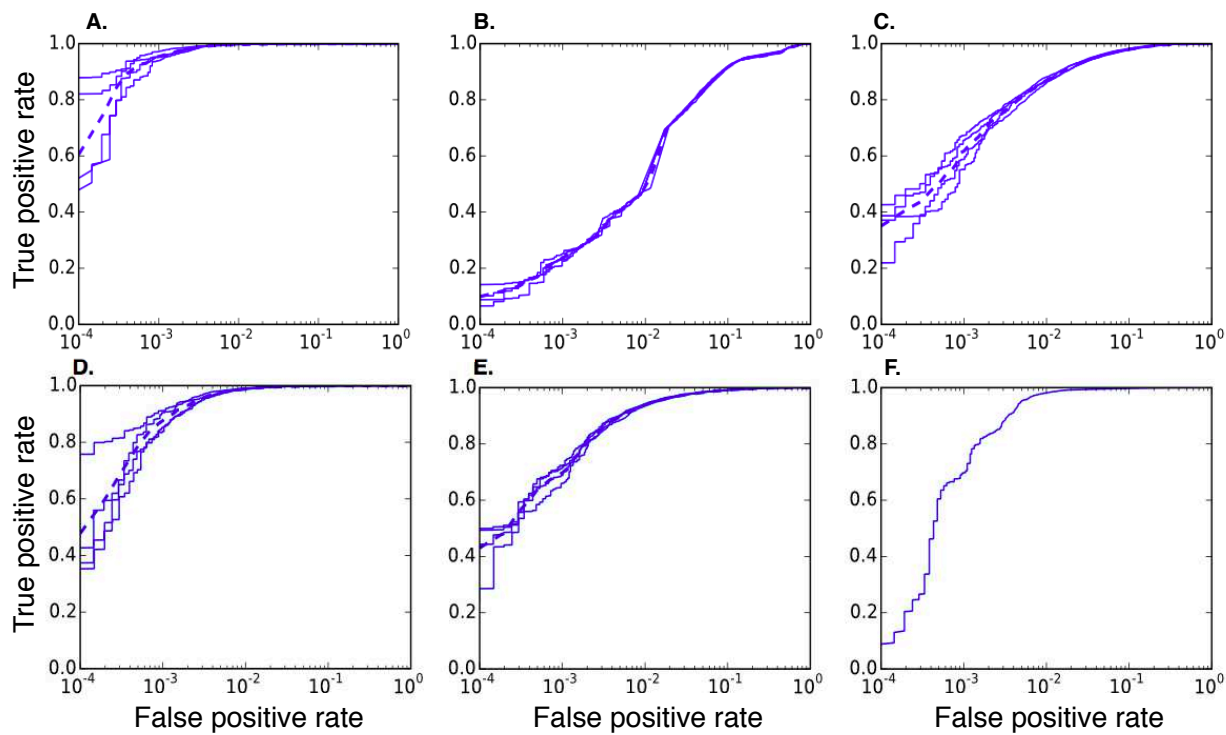}
   \caption{Six experiments showing the accuracy of our approach for different combination of feature types. For each experiment we show a set of solid lines, which are the ROC of the individual cross-validation folds, and a dotted line is the averaged value of these ROC curves. A. all four feature types;  B. only PE import features; C. only byte/entropy features; D.  only metadata features; E. only string features; F. all features after we train only on the samples whose compile timestamp is before July 31st, 2014 and test on samples whose compile timestamp is after July 31st, 2014, excluding samples with blatantly forged compile timestamps.}
   	\label{fig:validation}
\end{figure*}

\begin{figure}[h]
	\center
	\includegraphics[width=0.45\textwidth]{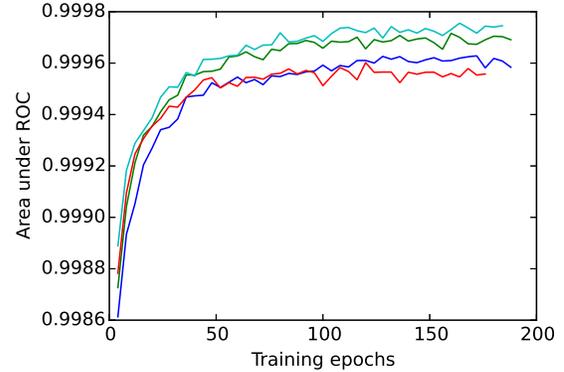}
   \caption{Plot showing the performance of our full 1024 feature model, as a function of training epochs, for each split of our cross-validation experiment.}
   	\label{fig:training_time}
\end{figure}

The first set of cross-validation experiments measured our system's individual accuracy of each of the four feature types described in Section \ref{sec:features}.  For these experiments we reduced the size of the neural network input layer to 256 and training our network for 200 epochs or until training error falls below $0.02$ for each fold.  Our fifth experiment was conducted in the same way, but included all of our features and used the 1024 neurons input layer.
The results of these validations are shown in Fig. \ref{fig:validation} A, B,C, D, and E, and is also summarized in Table \ref{tab:r1}. These validation results show there is significant variation in how each of our feature sets performs.  Using all of the features together produced the best results, with an average of 95.2\% of malicious binaries not seen in training detected at a 0.1\% false positive rate, with area under the roc (AUC) of 0.99964. Fig. \ref{fig:training_time} shows that our ROC improves with the number of epochs, and we are not suffering from overfitting. For our full dataset, it takes approximately 15 seconds to train one epoch using a single GPU, so the full model can be effectively trained in about 40 minutes.

\begin{table}
\centering
\caption{Estimated TPR at 0.1\% FPR and AUC, for the corresponding plots in Fig. \ref{fig:validation}}
\begin{tabular}{| l |  l | l |} \hline
Features & TPR & AUC \\ \hline
A. All   & 95.2\%  & 0.99964  \\  \hline
B. PE Import  & 22.8\%  &  0.95785  \\  \hline
C. Byte/entropy  & 61.1\%  & 0.99145  \\  \hline
D. Metadata  & 86.7\%  &  0.99912  \\  \hline
E. Strings  & 68.8\%  & 0.99581  \\  \hline
F. All (Time Split)  &  67.7\%  &  0.99794 \\
\hline\end{tabular}
\label{tab:r1}
\end{table}

In terms of independent feature sets, our PE metadata features perform best, with close to 87\% of malware binaries unseen in training detected at a 0.1\% false positive rate.  Our string features also perform quite well, with 69\% of unseen malware detected at a 0.1\% false positive rate.  While our byte-entropy features and import features don't perform as well as our PE metadata and string features, we found that they boost accuracy beyond what string and PE metadata features can provide on their own.

\subsection{Expected Deployment Performance}

One important question, if our classifier is to be deployed, is how to relate the cross-validated ROC to the expected performance in enterprise setting. To estimate expected performance, we observed the number of previously unseen binaries that were detected for the entire set of customers during a span of a few days. This gave us an expected average of around 5 previously unseen executed binaries per endpoint, per day. For FPR of 0.1\%, this would result in about five false positives per day, per 1000 endpoints, assuming our ROC curve is an accurate estimate of actual performance. We have confirmed this result by directly running our sensor on incoming data (endpoint binaries and Jotti stream) over several days, which yielded a similar performance estimate. Interestingly, some of the top false positives were anti-virus installers and dubious tools used by Invincea's product development team.

\subsection{Time Split Experiment}

A shortcoming of the standard cross-validation experiments is that they do not separate our ability to detect slightly modified malware from our ability to detect new malware toolkits and new versions of existing malware toolkits.  That is because most unique malware binaries, as defined by cryptographic checksums, are just automatically generated copies of various metamorphic malware designed to help it evade signature-based detection. 

Thus, to test our system's ability to detect genuinely new malware toolkits, or new malware versions, we ran a time split experiment that better estimates our system's ability to detect new malware toolkits and new versions of existing malware toolkits. We first extracted the compile timestamp field from each binary executable in our dataset.  Next we excluded binaries that had compile timestamps after July 31st , 2015 (the date on which the experiment was run) and binaries with compile timestamps before January 1st, 2000, since for those malware samples the authors have blatantly tampered with malware binaries' compile timestamps or the file is corrupted.  Finally, we split our test data into two sets: a set of binaries with compile timestamps before July 31st, 2014, and the set of binaries on or after July 31st, 2014.  Then, we trained our neural network (using all of the features described above) on the earlier dataset and tested on the later dataset. While we cannot completely trust that malware authors do not often modify the compile timestamps on their binaries, there is little motivation for doing so, and the distribution of dates matches what we know about our dataset sources, supporting that hypothesis.

The results of this experiment, as shown in Fig. \ref{fig:validation}F, demonstrate that our system performs substantially worse on this test, detecting 67.7\% of malware at a 0.1\% FPR, and approaching a 100\% detection rate at a 1\% FPR. The substantial degradation in performance is not surprising given the difficulty of detecting genuinely novel malware programs versus detecting malware samples that are new instances of known malicious platforms and toolkits. This result suggests that the classifier should be updated often using new data in order to main it's classification accuracy. This, however, can be done rapidly and cheaply for a neural network classifier.

\section{Related Work}
\label{sec:background}

Malware detection has evolved over the past several years, due to the increasingly growing threat posed by malware to large corporations and governmental agencies. Traditionally, the two major approaches for malware detection can be roughly split based on the approach that is used to analyze the malware, either static and dynamic analysis (see review \cite{egele2012survey}). In static analysis the malware file, or set of files, are either directly analyzed in binary form, or additionally unpacked and/or decompiled into assembly representation. In dynamic analysis, the binary files are executed, and the actions are recorded through hooking or some access into internals of the virtualization environment.

In principle, dynamic detection can provide direct observation of malware action, is less vulnerable to obfuscation \cite{moser2007limits}, and makes it harder to recycle existing malware. However, in practice, automated execution of software is difficult, since malware can detect if it is running in a sandbox, and prevent itself from performing malicious behavior. This resulted in an arms race between dynamic behavior detectors using a sandbox and malware \cite{anubis2015,fleck2013pytrigger}. Further, in a significant number of cases, malware simply does not execute properly, due to missing dependencies or unexpected system configuration. These issues make it difficult to collect a large clean dataset of malware behavior.

Static analysis, on the other hand, while vulnerable to obfuscation, does not require elaborate or expensive setup for collection, and very large datasets can be created by simply aggregating the binaries files. Accurate labels can be computed for all these files using anti-virus aggregator sites like VirusTotal \cite{virtualbox2015}. 

This makes static analysis very amenable for machine learning approaches, which tends to perform better as data size increases \cite{banko2001scaling}. Machine learning has been applied to malware detection at least since \cite{kephart1995biologically}, with numerous approaches since (see reviews \cite{egele2012survey,gandotra2014malware}). Machine learning consists of two parts, the feature engineering, where the author transforms the input binary into a set of features, and a learning algorithm, which builds a classifier using these features. 

Numerous static features have been proposed for extracting features from binaries: printable strings \cite{schultz2001data}, import tables \cite{weber2002toolkit}, byte $n$-grams \cite{abou2004n}, opcodes \cite{weber2002toolkit}, informational entropy \cite{weber2002toolkit}. Assortment of features have also been suggested during the Kaggle Microsoft Malware Classification Challenge \cite{kaggle2015}, such as opcode images, various decompiled assembly features, and aggregate statistics. However, we are not aware of any published methods that break the file into subsamples (\eg, using sliding windows), and creates a histogram of all the file's subsamples based on two or more properties of the individual subsample.

Potentially the feature space can become large, in those cases methods like locality-sensitive hashing \cite{indyk1998approximate,bayer2009scalable}, feature hashing (aka hashing ``trick'') \cite{weinberger2009feature, jang2011bitshred}, or random projections \cite{johnson1984extensions,indyk1998approximate,fradkin2003experiments,dahl2013large} have been used in malware detection. 

The large number of features, even after dimensionality reduction, can cause scalability issues for some machine learning algorithms. For example, non-linear SVM kernels require $O(N^2)$ multiplication during each iteration of the optimizer, where $N$ is the number of samples \cite{friedman2001elements}. $k$-Nearest Neighbors ($k$-NN) requires finding $k$ closest neighbors in a potentially large database of high dimensional data, during prediction, which requires significant computation and storage of all label samples. 

One popular alternative to the above are ensemble of trees (boosted trees or bagged trees), which can scale fairly efficiently by subsampling the feature space during each iterations \cite{breiman2001random}. Decision trees can adapt well to various data types, and are resilient to heterogeneous scales of values in feature vectors, so they exhibit good performance even without some type of data standardization. However, standard implementations typically do not allow incremental learning, and fitting the full dataset with large number of features could potentially require expensive hardware.

Recently, neural networks have emerged as a scalable alternative to the standard machine learning approaches, due to significant advances in training algorithms \cite{wozniak2015review,lecun2015deep}. Neural networks have been previously used in malware detection \cite{kephart1995biologically,dahl2013large,benchea2014combining}, though it is not clear how to compare results, since datasets are different, in addition to the various pre-filtering of samples that is done before evaluation.

\section{Conclusion}
\label{sec:conclusion}

In this paper we introduced a deep learning based malware detection approach that achieves a detection rate of 95\% and a false positive rate of 0.1\% over an experimental dataset of over 400,000 software binaries.  Additionally, we have shown that our approach requires modest computation to perform feature extraction and that it can achieve good accuracy over our corpus on a single GPU within modest timeframes.  

We believe that the layered approach of deep neural networks and our two dimensional histogram features provide an implicit categorization of binary types, allowing us to directly train on all the binaries, without separating them based on internal features, like packer types, and so on. 

Neural networks also have several properties that make them good candidates for malware detection. First, they can allow incremental learning, thus, not only can they be training in batches, but they can re-trained efficiently (even on an hourly or daily basis), as new training data is collected. Second, they allow us to combine labeled and unlabeled data, through pre-training of individual layers \cite{hinton2006fast}. Third, the classifiers are very compact, so prediction can be done very quickly using low amounts of memory. 

Further attesting to the value of our approach, our system has proven a crucial part of our company's overall malware detection and prevention product and has been deployed to our cloud security analytics platform, which is currently performing detection on files streaming from thousands of customer endpoints.

\section{Software and Data}

The feature extraction code, data matrix, the label vector, and the neural network code is available at \url{https://github.com/konstantinberlin/Malware-Detection-Using-Two-Dimensional-Binary-Program-Features}.

\section{Acknowledgement}
We would like to thank Aaron Liu at Invincea Inc. for providing crucial engineering support over the course of this research.  We also thank the Invincea Labs data science team, including Alex Long, David Slater, Giacomo Bergamo, James Gentile, Matt Johnson, and Robert Gove, for their feedback as this work progressed.

\bibliographystyle{abbrv}
\bibliography{static_detection}
\end{document}